\documentclass[conference]{IEEEtran}
\IEEEoverridecommandlockouts
\usepackage{cite}
\usepackage{amsmath,amssymb,amsfonts}
\usepackage{algorithmic}
\usepackage{graphicx}
\usepackage{textcomp}
\usepackage{xcolor}
\usepackage{color,soul}
\usepackage{bm}
\usepackage{subfig}
\usepackage{float}

\def\BibTeX{{\rm B\kern-.05em{\sc i\kern-.025em b}\kern-.08em
    T\kern-.1667em\lower.7ex\hbox{E}\kern-.125emX}}

\DeclareMathOperator{\EX}{\mathbb{E}}% expected value
\begin{document}

\title{A Measurement-Based Robust Non-Gaussian Process Emulator Applied to Data-Driven Stochastic Power Flow\\

\thanks{The authors gratefully acknowledge the financial support of NSF via the grant ID 1917308.}
}

\author{\IEEEauthorblockN{Pooja Algikar}
\IEEEauthorblockA{\textit{Electrical and Computer Engineering}\\
\textit{Northern Virginia Center, Virginia Tech}\\
Falls Church, VA 22043 \\
apooja19@vt.edu}
\and
\IEEEauthorblockN {Yijun Xu}
\IEEEauthorblockA{\textit{Electrical and Computer Engineering} \\
\textit{Northern Virginia Center, Virginia Tech}\\
Falls Church, VA 22043 \\
yijunxu@vt.edu}
\and
\IEEEauthorblockN{Lamine Mili}
\IEEEauthorblockA{\textit{Electrical and Computer Engineering} \\
\textit{Northern Virginia Center, Virginia Tech}\\
Falls Church, VA 22043\\
lmili@vt.edu}

}

\maketitle

\begin{abstract}
In this paper, we propose a robust non-Gaussian process emulator based on the Schweppe-type generalized maximum likelihood estimator, which is trained on metered time series of voltage phasors and power injections to perform stochastic power flow. Power system data are often corrupted with outliers caused by fault conditions, power outages, and extreme weather, to name a few. The proposed emulator bounds the influence of the outliers using weights calculated based on projection statistics, which are robust distances of the data points associated with the rows vectors of the factor space. Specifically, the developed estimator is robust to vertical outliers and bad leverage points while retaining good leverage points in the measurements of the training dataset. The proposed method is demonstrated on an unbalanced radial IEEE 33-Bus system heavily integrated with renewable energy sources.
\end{abstract}


\section{Introduction}
Modern power systems are rapidly transitioning towards net-zero carbon emissions, now more than ever. For a reliable operation and control under heavy penetration of renewable energy sources (RES) and distributed generations (DGs), advanced sensing and metering devices are proliferating, thereby enhancing the observability of the previously passive networks. Alongside RES and DGs, electrified heat and transport systems introduce stochastic dynamics in the power distribution systems.

Fine-grained data are available via an increasing number of measurement equipment such as SCADAs, intelligent electronic devices (IEDs), smart meters, and phasor measurement units (PMUs), among others, which are installed on the power transmission and distribution systems. 
In the United States, photovoltaic (PV) generation has grown from $3,063$ GW to $116,692$ GW while  wind generation (WG) has increased from $95,148$ GW to $341,416$ GW in the last decade. With this unprecedented growth of RES that are non-deterministic in nature, the steady-state analysis of power systems using the stochastic power flow equations to analyze their dynamics is of paramount importance for operational control actions. 
In the literature, various methodologies were proposed to solve the probabilistic power flow with or without the consideration of correlation among the renewable generations \cite{Lin2018ComparisonGeneration}. 
Among them, the Monte Carlo (MC) method is the most well-known one for its flexibility in the implementation; however, it is computationally expensive as one needs to perform thousands of simulation runs to assess the meaningful statistical properties of nodal voltages and line power flows. Hence, it comes with no surprise that the emulator as a surrogate approach is gaining increase attention today for its computational efficiency compared with the MC approach. Compared with other analytical approaches that typically rely on the linear and Gaussian assumptions, this surrogate approach is \emph{derivative-free} and not limited to any specific probability distribution while considering  a nonlinear power system model.  Interestingly, a large-scale  power system model is typically reduced-modeled as a simple surrogate using just a small number of power flow simulator runs \cite{Rocchetta2020AGrids}, \cite{Rocchetta2018ADeficiency}, \cite{Xu2020ProbabilisticEmulator}. Consequently, the computational efficiency of the probabilistic power flow analysis can be greatly improved by using a well-trained emulator. Typical examples for the surrogate methods include the parametric polynomial-chaos surrogate \cite{ren2015probabilistic} and the non-parametric Gaussian process emulator surrogate \cite{Xu2020ProbabilisticEmulator}.

One weakness of the current application of the emulator is that a given probability distribution is assumed, typically the Gaussian distribution for the load, the Weibull distribution for the wind speed, and the Beta distribution for the solar, among others. However, in practice, the assumed  probability distribution may be a poor approximation of the actual one, yielding  inaccurate results. Therefore, some papers in the literature propose to improve the surrogate approaches by using raw data. For example, Wang ~\emph{et al.} \cite{wang2020data} developed a data-driven polynomial chaos approach to directly estimate the statistics from the data while  Xu~\emph{et al.} proposed to avoid the parametric distribution for Gaussian process emulator via a fully non-parametric approach \cite{xu2020data}. However, all these methods are relying on raw data without considering outliers. This is indeed an over-bold assumption that does not hold in practice. It is well known that the WG time series data are frequently corrupted by communication errors, wind turbine outages, and curtailment \cite{Ye2016IdentificationData} while PV time series data are corrupted by signal noise, sensor failure, communication equipment failure, maximum power tracking abnormalities, array shutdown, and power limitation, to name a few \cite{Li2020OutlierApplication}. So, it comes as no surprise that several papers have addressed the outlier issues in the raw data of the RES. For instance, an image-processing-based method is advocated by Long
~\emph{et al.} in \cite{long2019image}. Similarly, a data-mining-based approach is proposed by Zheng ~\emph{et al.} in \cite{zheng2014raw}. However, the data-driven surrogate approach will yield strongly biased uncertainty quantification if it fails to suppress outliers in the raw data.

To address this problem, we develop in this paper a robust statistical technique for real-time data-driven probabilistic power flow analysis based on the non-Gaussian process emulator (NGPE). The proposed NGPE is based on the Schweppe-type generalized maximum likelihood estimator (SHGM) that can handle up to $25\%$ of outliers in the training data. Recall that outliers may be either vertical outliers or bad leverage points. In power systems, leverage points are power flows on relatively short lines or power injection on buses with relatively many incident lines.
Gross errors on them yield the so-called bad leverage points. Outliers on the other measurements are vertical outliers \cite{Mill1996robustStatistics}. 
 The robustness of the NGPE is demonstrated on the radial IEEE-33 bus system for each type of outlier. 

\section{Non-Gaussian Process Emulator}
\subsection{Formulation of the Stochastic Power Flow in the Gaussian Process Emulator Framework}
Let us consider a set of power flow equations described by
\begin{equation}
    \mathbf{y}=\mathbf{F}(\mathbf{x})+\bm{\epsilon},
\end{equation}
where the $i^{th}$ realization of input, $\mathbf{x}\in \mathbb{R}^{2p}$, is given by $\bm{x}=[\mathbf{x}_1,\mathbf{x}_2, \hdots, \mathbf{x}_n]^{T}$ are the active and reactive power injection measurements of all the buses at time $\bm{T}=[{t}_1,{t}_2, \hdots, {t}_n]^{T}$. The $f(\mathbf{x}): \mathbb{R}^{2p}\rightarrow\mathbb{R}$ can be considered as a power flow model whose output $\mathbf{y}$ is a voltage magnitude or phase angle measurement at a Bus with an independent and identically distributed measurement noise ${\epsilon}\sim\mathcal{N}(0,\sigma^{2})$. The model output at each instance is a realization of the Gaussian process following a joint multivariate normal probability distribution given by
\small
\begin{equation}\label{eq62}
    \begin{bmatrix}
        f(\mathbf{x}_1)\\
        \vdots\\
        f(\mathbf{x}_n)\\
    \end{bmatrix}\sim \mathcal{N}\left( \begin{bmatrix} m(\mathbf{x}_1)\\
        \vdots\\
        m(\mathbf{x}_n)\\    \end{bmatrix},\begin{bmatrix}
            k(\mathbf{x}_1,\mathbf{x}_1) & \dots & k(\mathbf{x}_1,\mathbf{x}_n)\\
            \vdots &       &     \vdots\\
            k(\mathbf{x}_n,\mathbf{x}_1) & \dots & k(\mathbf{x}_n,\mathbf{x}_n)  
        \end{bmatrix}\right).
\end{equation}
\normalsize
Here, the mean function is $\bm{m}(\mathbf{x})=\mathbf{H}(\mathbf{x})\bm{\beta}$ and $\bm{k}(\mathbf{x},\mathbf{x})=\textrm{Cov}(\mathbf{y}(\mathbf{x}),\mathbf{y}(\mathbf{x}))$ is a kernel function  that represents a covariance function, where $\mathbf{H}(\mathbf{x}):\mathbb{R}^{p}\rightarrow\mathbb{R}^{q}$ denotes the basis function that can be chosen to model the belief about the degree of non-linearity of the power system. To capture the non-linearity of the power flow equation, we consider the quadratic basis function given by $\mathbf{H}(\mathbf{x})=[\bm{1},\bm{x}_{1},\hdots\mathbf{x}_{p},\mathbf{x}_{1}^2,\hdots,\mathbf{x}_{p}^2]$. A stationary squared exponential kernel function  $\bm{k}(\mathbf{x},\mathbf{x})=\tau^2\bm{R}(\mathbf{x},\mathbf{x})$ where $\bm{R}(\cdot)$ models the correlation between two output data points given by
\begin{equation}
R(\mathbf{x}_{i},\mathbf{x}_{j}|\bm{s})=\textrm{exp}\left(-\sum_{k=1}^{p}\frac{(\textrm{x}_{ik}-\textrm{x}_{jk})^2}{{l}_k^2}\right). 
\end{equation}
The outputs $\mathbf{y}=[y_1,y_2,\hdots,y_n]^{T}$ follow a multivariate Gaussian distribution with the covariance matrix $\bm{\Sigma}(\mathbf{X})=\bm{k}(\mathbf{X},\mathbf{X})+\sigma^{2}_{n} I_n$, where $\sigma^{2}_{n}$ stands for variance of the diagonally added noise element in the measurements typically known as "nugget" with zero mean which accounts for model uncertainty and numerical stability, as follows:
\begin{equation}
    \mathbf{y}|\mathbf{X}\sim\mathcal{N}\left(\bm{m}(\mathbf{X}),\bm{\Sigma}(\mathbf{X})\right).
\end{equation}
Suppose that we are interested in the output of the model $f(\cdot)$ at the $t+N^{*}$ instance for which  the associated test point in the input feature space is denoted by $\mathbf{x}^*\in \mathbb{R}^p$. The model output $y^{*}$ at a test point combined with the training inputs follow a joint multivariate Gaussian distribution given by
\begin{equation}\label{eq5}
    \begin{bmatrix}
      \mathbf{Y}\\
     \mathbf{y}^{*}
    \end{bmatrix}\sim \mathcal{N}\left( \begin{bmatrix}
      \bm{m}(\mathbf{X})\\
      \bm{m}(\mathbf{x}^*)\\
    \end{bmatrix},\begin{bmatrix}
    \bm{\Sigma}({\mathbf{X}})& \mathbf{C}(\mathbf{x}^{*})\\
        \mathbf{C}^{T}(\mathbf{x}^{*})&\mathbf{V}(\mathbf{x}^{*})\\
    \end{bmatrix} \right),
\end{equation}
 where $\mathbf{C}(\mathbf{x}^{*})=\bm{k}(\mathbf{X},\mathbf{x}^{*}),\mathbf{C}^{T}(\mathbf{x}^{*})=\bm{k}\mathbf{(x^{*},X)}$ and $\mathbf{V}(\mathbf{x}^{*})=\bm{k}(\mathbf{x}^{*},\mathbf{x}^{*})$. The a priori probability distribution assumed for the simulator output  ${f}(\mathbf{x}^{*})|\mathbf{x}^{*}$ is given by
 \begin{equation}\label{eq8}
    {f}(\mathbf{x}^{*})|\mathbf{x}^{*}\sim\mathcal{N}\left( \bm{m}(\mathbf{x}^*),\bm{k}\mathbf{({x}^{*},{x}^{*})}\right).
    \end{equation}
Using the Bayes theorem, we infer the posterior distribution of the quantities that are of interest to us, which are ${f}(\mathbf{x}^{*})$, conditioned upon the training points $(\mathbf{y},\mathbf{X})$. Formally, we have  
\begin{equation}\label{eq9}
  {f}(\mathbf{x}^{*})|\mathbf{x}^{*},\mathbf{y},\mathbf{X}\sim\mathcal{N}\left(\bm{\mu}^{*}(\mathbf{X}),\bm{\Sigma}^{*}(\mathbf{X})\right).
\end{equation}
Deriving the conditional distribution corresponding to \eqref{eq5}, we get the predictive equations as follows:
\begin{equation}\label{eq10}
 \bm{\mu}^{*}(\mathbf{x^{*}})
  = \bm{m}(\mathbf{x}^{*})+\mathbf{C}^{T}(\mathbf{x}^{*})\mathbf{R}^{-1}\mathbf{r},
\end{equation} 
\begin{equation}\label{eq11}
    \bm{\Sigma}^{*}(\mathbf{x^{*}})=\mathbf{V}(\mathbf{x}^{*})-\mathbf{C}^{T}(\mathbf{x}^{*})\mathbf{R}^{-1}\mathbf{C}(\mathbf{x}^{*}),
\end{equation}
where $\bm{r}=\mathbf{y}-\mathbf{H}(\mathbf{X})\bm{\beta}$. The mean function given by \eqref{eq10} acts as a surrogate model that very closely captures the behavior of the power flow and the variance of the model given by \eqref{eq11} quantifies the associated uncertainty.  

\subsection{Robust Estimation of the Regression Weight Vector}
The regression weight vector $\bm{\beta}$ is estimated by solving the regression model of the mean function with reference to a realization of the GP emulator, the latter of which is given as
\begin{equation}
    \mathbf{y}(\mathbf{X})=\mathbf{H}(\mathbf{X})\bm{\beta}+\bm{e},
\end{equation}
where $\EX(\bm{e})=0$ and $\textrm{Cov}(\bm{e})=\bm{\Sigma}(\mathbf{X})$.
We now estimate the $\bm{\beta}$ in a robust manner using the SHGM estimator.  The SHGM estimator minimizes an objective function given by
\begin{equation}\label{eq14}
    J(\bm{\beta})=\underset{\hat{\bm{\beta}}}{\mathrm{min}}\sum_{i=1}^{n} w_i^2\rho\bigg(\frac{r_i}{w_i s}\bigg),
\end{equation}
where $\rho(\cdot)$ is a non-linear function of the standardized residuals, $r_{Si}=\frac{r_i}{w_i s}$, defined as

 \begin{equation}\label{13}
    \rho(r)=\begin{cases}
    \frac{r^2}{2}     &\text{for } r< b,\\
    br-\frac{b^2}{2}  &\text{for } r\geq b.
\end{cases}
\end{equation}
We choose the Huber $\rho-$function because of its convexity and its quadratic characteristic at its center and because it yields an estimator with a bounded total influence function.
As for the weight $w_i$ associated with the $i^{th}$ residual  $r_i= y_i-\bm{h}^{T}(\mathbf{x}_{i})\hat{\bm{\beta}}$, it is calculated based on the projection statistic $PS_i$. It is a function of the  outlyingness of the associated data point among the point cloud defined by the row vectors of the basis function. It is small for the points having large outlyingness. Recall that the ${PS_i}$ is the maximum of the standardized projection distances obtained by projecting the point cloud on the directions that originate from the coordinate-wise median and that passe through each of the data points \cite{Mill1996robustStatistics}). Formally, we have 
 \begin{equation}\label{eq36}
     {PS}_{i}=\underset{||\bm{u}||=1}{max}\; \frac{\mathbf{h}_{i}^{T}\bm{u}-\underset{j}{med}(\mathbf{h}_{j}^{T}\bm{u})}{1.4826\;\underset{k}{med}\;|\mathbf{h}_{k}^{T}\bm{u}-\underset{j}{med}(\mathbf{h}_{j}^{T}\bm{u})|} .
 \end{equation}
The weights are then calculated as
\begin{equation}\label{15}
   w(\bm{h}_i)=\begin{cases}
    1,& \text{for $\bm{h}_i$ not a leverage point}\\
    br_{i}-\frac{b}{PS_i^2}, & \text{for $\bm{h}_i$ a leverage point}
\end{cases}
\end{equation}
where $b=\chi^{2}{(\nu,0,975)}$ is defined as the threshold for the outlier identification where $\nu$ is the total number of non-zero elements in the basis function row. 
They downweight the bad leverage point and vertical outliers while retaining the good leverage points.

The SHGM estimator is a solution of 
\begin{equation}\label{14}
    \sum_{i=1}^{m}\bm{w}_{i}\mathbf{h}_{i}\Psi(\bm{r}_{Si})=0,
\end{equation} 
where $\Psi(\bm{r}_{Si})=\frac{\partial\bm{\rho}(\bm{r}_{Si})}{\partial\bm{r}_{Si}}$.
This equation is solved using the iteratively reweighted least-squares (IRLS) algorithm as follows: \eqref{14} can be re-written in matrix form as
\begin{equation}\label{2}
\sum_{i=1}^{m}\bm{q}\bigg(\frac{\bm{r}_i}{\bm{w}_i s}\bigg)\mathbf{h}_i\bm{r}_i=\bm{0},
\end{equation}
where $\bm{q}({\bm{r}_{Si}})=\frac{\Psi(\bm{r}_{Si})}{\bm{r}_{Si}}$.  For the case of Huber $\rho$-function, ${q}$-function is defined as
\begin{equation}\label{1}
   {\bm{q}}(\bm{r}_{S_i})=\begin{cases}
    1,&  \bm{r}_i\leq c\\
    \frac{b\; \textrm{sign}(\bm{r}_{S_i})}{\bm{r}_{Si}}, & \textrm{otherwise}\\
\end{cases}.
\end{equation}
Finally, substituting the expression for $\bm{r}$ and $\mathbf{Q}=\textrm{diag}(\bm{q}(\bm{r}_{Si}))$ in \eqref{2}, we get
\begin{equation}
     \mathbf{H}^T\mathbf{Q}(\bm{y}-\mathbf{H}\bm{\beta})=\mathbf{0}.
\end{equation}
Since $\mathbf{Q}$ is a function of $\bm{\beta}$, we need to iteratively solve for $\bm{\beta}$ as follows:
\begin{equation}\label{eq19}
    \bm{\beta}^{i+1}=(\mathbf{H}^T\mathbf{Q}^i \mathbf{\Sigma}^{-1}\mathbf{H})^{-1}\mathbf{H}^T\mathbf{Q}^i \mathbf{\Sigma}^{-1} \mathbf{y}.
\end{equation}
One interesting feature of the SHGM estimator is that it reduces to (a) the generalized least absolute value estimator and (b) the weighted least squares estimator as as $c$ tends zero and to infinity, respectively. As a result, if $c$ is too small, there is a risk of good leverage points being severely downweighed and if it is too large, the bias of the estimates though bounded may be significant. Therefore, $c$ is typically chosen to be equal to $1.5$.  In addition, this threshold value guarantees a high statistical efficiency of the estimator at Gaussian distribution, namely $95\%$.   
\subsection{Hyperparameter Estimation}
In this subsection, the procedure for the robust estimation of hyperparameters ($\bm{l},\tau,\sigma^{2}_{n}$) of the NGPE associated with the covariance function is discussed.

With the observation set available from MC simulation of the code, $(\mathbf{y},\mathbf{X})$, we can estimate the
hyperparameters $\bm{\eta}$ in the GPE.
The MLE estimate of hyperparameters can be formulated as
\begin{equation}\label{eq76}
  (\widehat{\bm{l}},\widehat{\tau},\widehat{\sigma}_{n}^2)=\underset{\bm{l},\tau,\sigma^{2}_{n}}{\mathrm{arg\, max}} \, \textrm{log}\,  L\left(\mathbf{Y}|\mathbf{X},\widehat{\bm{\beta}},\bm{l},\tau,\sigma_{n}^2\right),
\end{equation}
Here, $\widehat{\bm{\beta}}$ represents the converged estimated weight vector expressed in \eqref{eq19}.  Simplifying $\textrm{log}\,L$ further, we get
\begin{multline}\label{eq77}
    \textrm{log}\,L\left(\mathbf{Y}|\mathbf{X},\widehat{\bm{\beta}},\bm{l},\tau,\sigma_{n}^2\right)\\=\
    -\frac{1}{2}(\mathbf{Y}-\mathbf{F}\widehat{\bm{\beta}})^T \left[\bm{k}(\mathbf{X},\mathbf{X}|\bm{l},\tau)+\sigma_{n}^2\mathbf{I}_n \right]^{-1}(\mathbf{Y}-\mathbf{F}\widehat{\bm{\beta}})\\-\frac{n}{2}\textrm{log}\, (2\pi)-\frac{1}{2}\textrm{log}\, |[\bm{k}(\mathbf{X},\mathbf{X}|\bm{l},\tau)+\sigma_{n}^2\mathbf{I}_n|.\\
 \end{multline}

Substituting the expression of $\widehat{\bm{\beta}}$ leads to
\begin{align}
    {\psi}(\bm{l},\tau,\sigma_{n}^2)= \textrm{log}\;|\bm{k}(\mathbf{X},\mathbf{X}|\bm{l},\tau)+\sigma_{n}^2\mathbf{I}_n|.
\end{align}
Then, the MLE estimate of $(\bm{l},\tau,\sigma_{n}^2)$ reduces to
\begin{equation}
    (\widehat{\bm{l}},\widehat{\tau},\widehat{\sigma}_{n}^2)= \underset{\bm{l},\tau,\sigma_{n}^2}{\mathrm{arg\, min}} \,(\bm{l},\tau,\sigma_{n}^2).
\end{equation}
With the estimated $(\widehat{\bm{l}},\widehat{\tau},\widehat{\sigma}_{n}^2)$, we can now re update $\widehat{\bm{\beta}}=\widehat{\bm{\beta}}(\widehat{\bm{l}},\widehat{\tau},\widehat{\sigma}_{n}^2)$.
We utilize a gradient-based
optimizer as described in \cite{gpmlbook}. The expression for the gradient of the reduced log-likelihood $\psi(\bm{l},\tau,\sigma_{n}^{2})$ with respect to $\sigma_n^2$ is given by    
\begin{equation}\label{eq83}
    \frac{\partial {\psi}(\bm{l},\tau,\sigma_{n}^2)}{\partial \sigma_{n}^2}=\textrm{trace}\left(\left(\bm{k}(\mathbf{X},\mathbf{X})+\sigma_{n}^2\mathbf{I}_n\right)^{-1}\right),
\end{equation}
and the derivative of the same with respect to $\bm{l}$ is derived as
\begin{equation}\label{eq84}
    \frac{\partial {\psi}(\bm{l},\tau,\sigma_{n}^2)}{\partial \bm{l}}=\left[\frac{\partial {\psi}(\bm{l},\tau,\sigma_{n}^2) }{\partial l_1}, \dots, \frac{\partial {\psi}(\bm{l},\tau,\sigma_{n}^2)}{\partial l_m} \right],
\end{equation}
where 
\small
\begin{equation}\label{eq86}
   \frac{\partial {\psi}(\bm{l},\tau,\sigma_{n}^2)}{\partial l } = (-2)\textrm{trace}\Biggl(\left(\bm{k}(\mathbf{X},\mathbf{X})+\sigma^2\mathbf{I}_n\right)^{-1}\bm{S}(l)\Biggr),
 \end{equation}
 \normalsize
 and
 \small
 \begin{equation}
    \bm{S}(l)=\begin{bmatrix}
       \frac{(\textrm{x}_{1l}-\textrm{x}_{1l})^2}{l^3}R(\mathbf{x}_1,\mathbf{x}_1) &\dots& \frac{(\textrm{x}_{1l}-\textrm{x}_{nl})^2}{l^3}R(\mathbf{x}_1,\mathbf{x}_n) \\
       \vdots & \ddots & \vdots\\
       \frac{(\textrm{x}_{nl}-\textrm{x}_{1l})^2}{l^3}R(\mathbf{x}_n,\mathbf{x}_1) & \dots & \frac{(\textrm{x}_{nl}-\textrm{x}_{nl})^2}{l^3}R(\mathbf{x}_n,\mathbf{x}_n)
   \end{bmatrix}
\end{equation}
 \normalsize
Also, the derivative with respect to $\tau$ is given by
\small
\begin{equation}\label{eq85}
  \frac{\partial {\psi}(\bm{l},\tau,\sigma_{n}^2)}{\partial \tau }=(2\tau) \textrm{trace}\Biggl(\left(\bm{k}(\mathbf{X},\mathbf{X})+\sigma^2\mathbf{I}_n\right)^{-1}{\bm{R}(\mathbf{X},\mathbf{X})}\Biggr),
\end{equation} 
\normalsize
where
\begin{equation}
    \bm{R}(\mathbf{X},\mathbf{X})=\begin{bmatrix} {R}(\mathbf{x}_1,\mathbf{x}_1) & \dots & {R}(\mathbf{x}_1,\mathbf{x}_n)\\ 
  \vdots & \ddots & \vdots\\
  {R}(\mathbf{x}_n,\mathbf{x}_1)& \dots & {R}(\mathbf{x}_n,\mathbf{x}_n).
 
  \end{bmatrix}
\end{equation}
\section{Simulation Results}
\begin{figure}%
    \centering
    \subfloat[\centering ]{{\includegraphics[height=2.5cm,width=9cm]{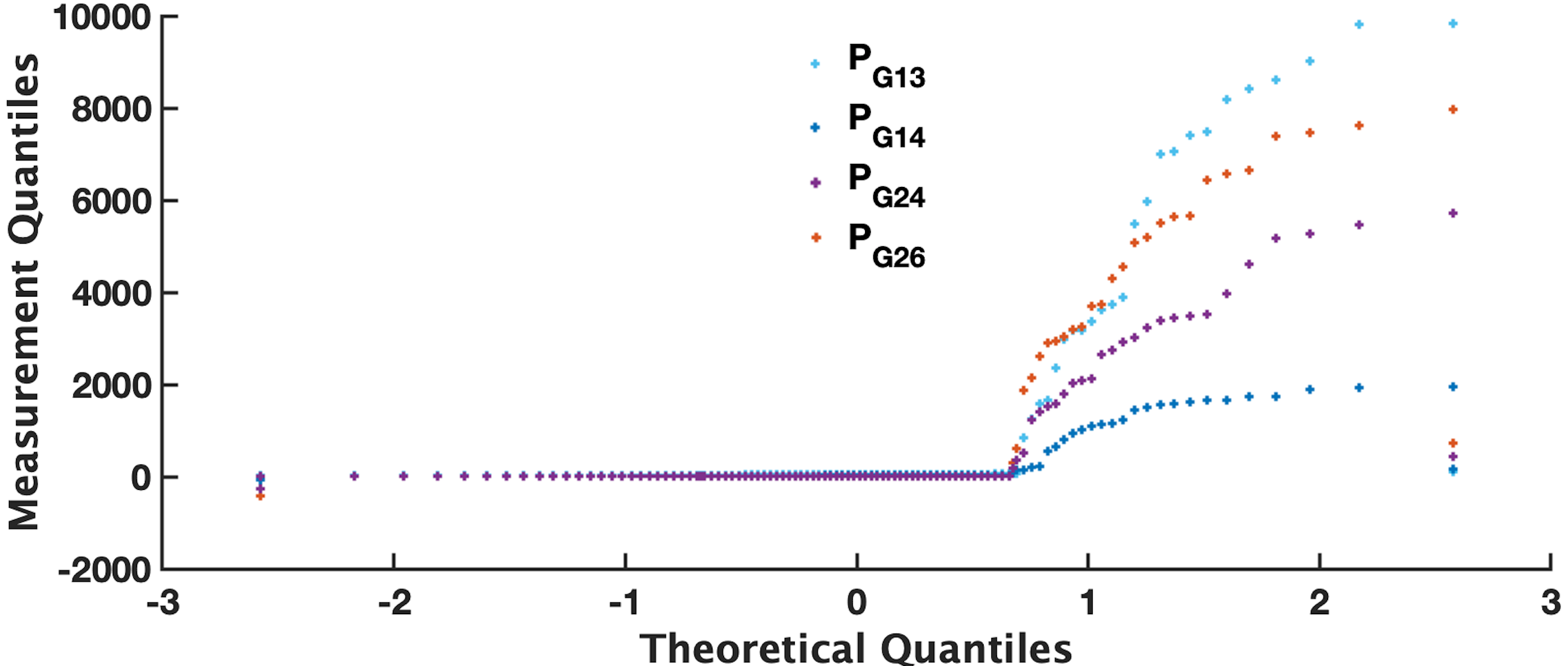} }}%
    \qquad
    \subfloat[\centering  ]{{\includegraphics[height=2.5cm,width=9cm]{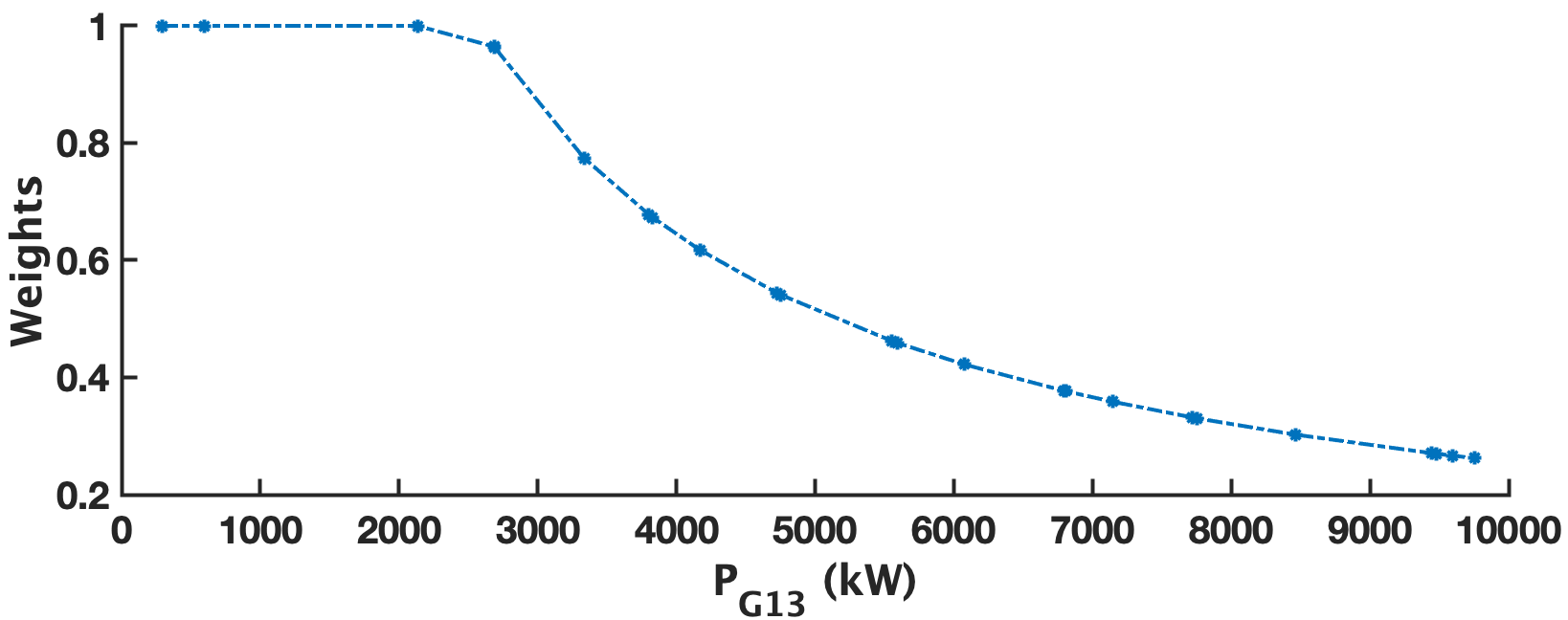}}}%
    \caption{Outliers added in the data; (a) the QQ-plot of the measurements with $25\%$ of added outliers; (b) the plot of weights of the SHGM estimator  vs magnitude  of outliers}%
    \label{DATA}
\end{figure}

\begin{figure}%
    \centering
    \subfloat[\centering ]{{\includegraphics[height=3cm,width=9cm]{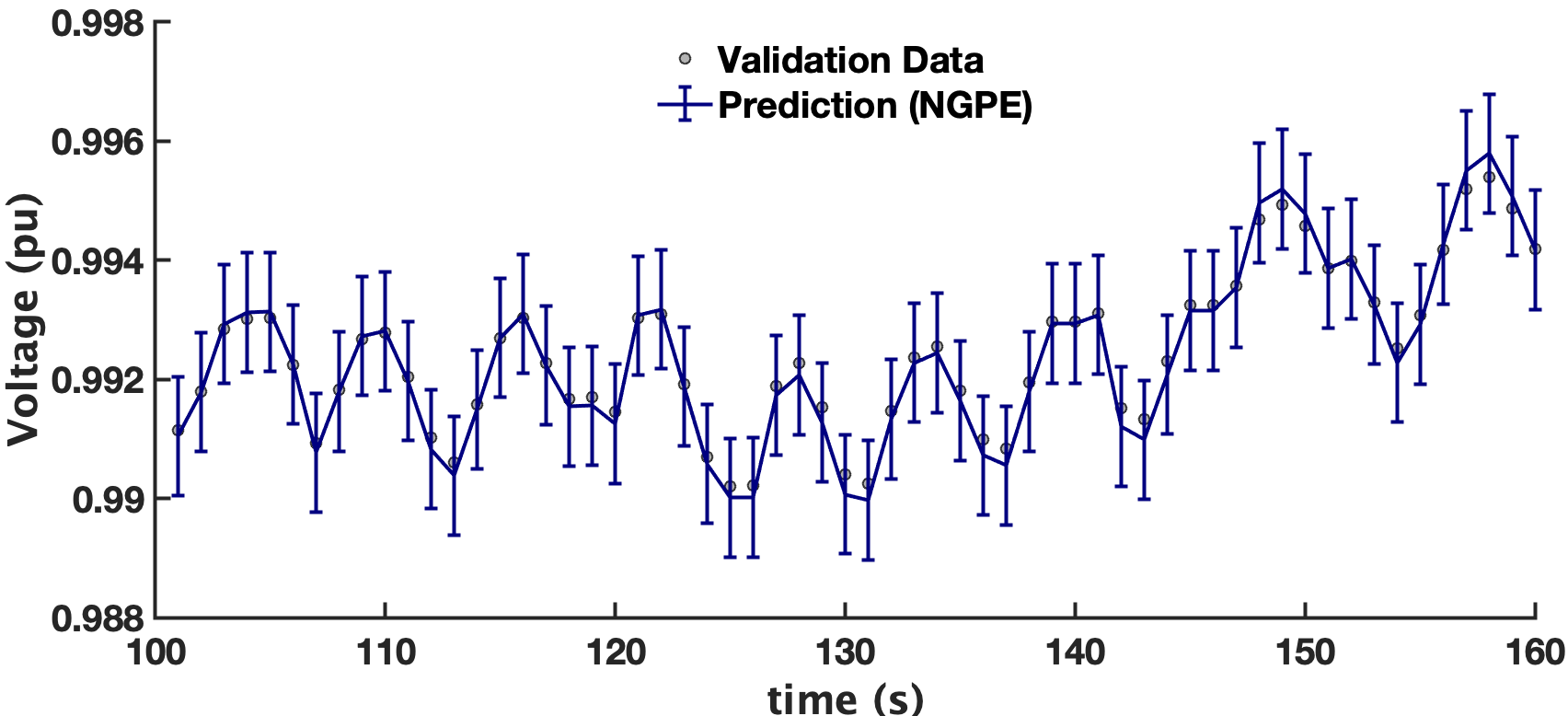} }}%
    \qquad
    \subfloat[\centering  ]{{\includegraphics[height=4.5cm,width=4.5cm]{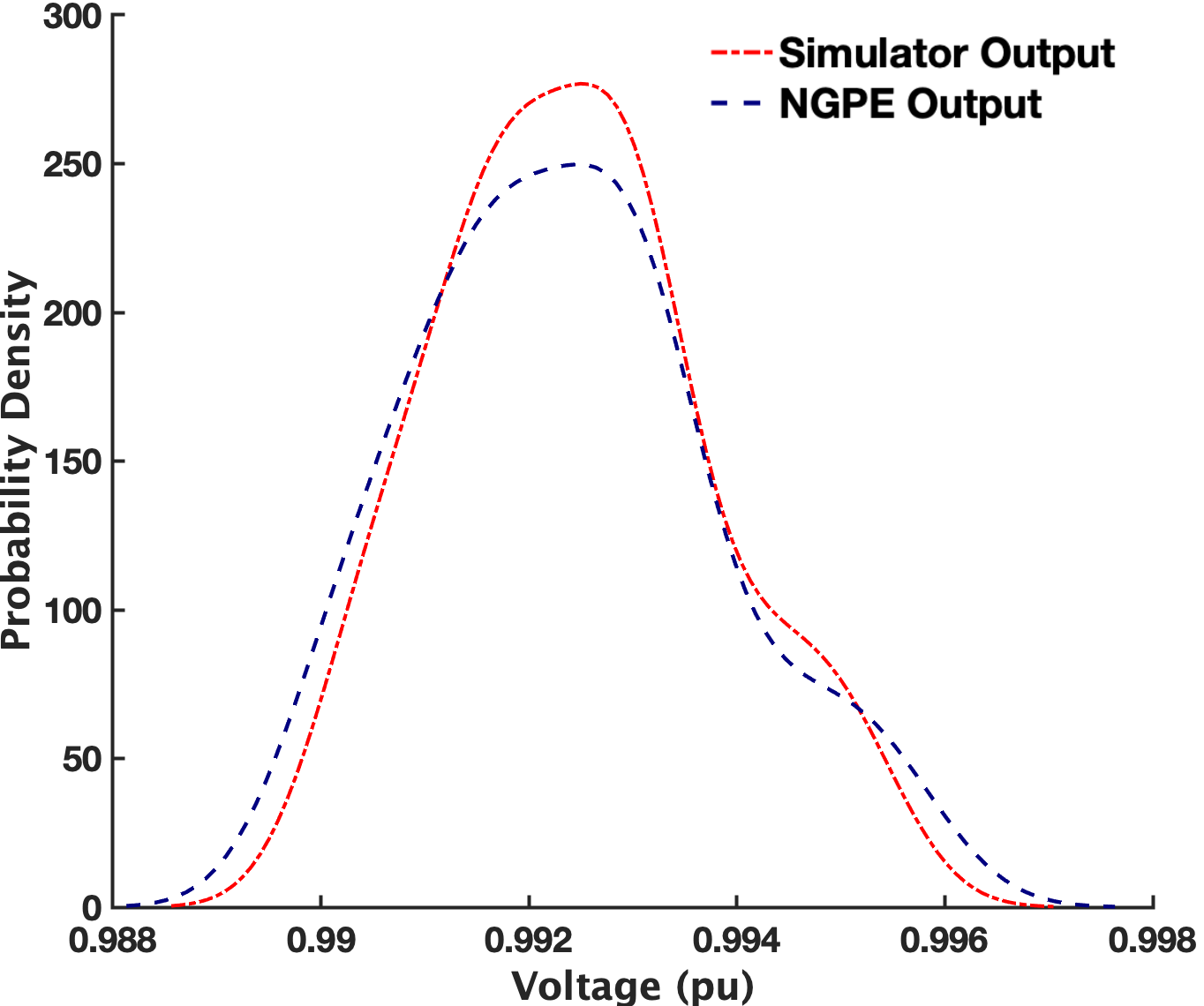} }}%
    \subfloat[\centering ]{{\includegraphics[height=4.5cm,width=4.5cm]{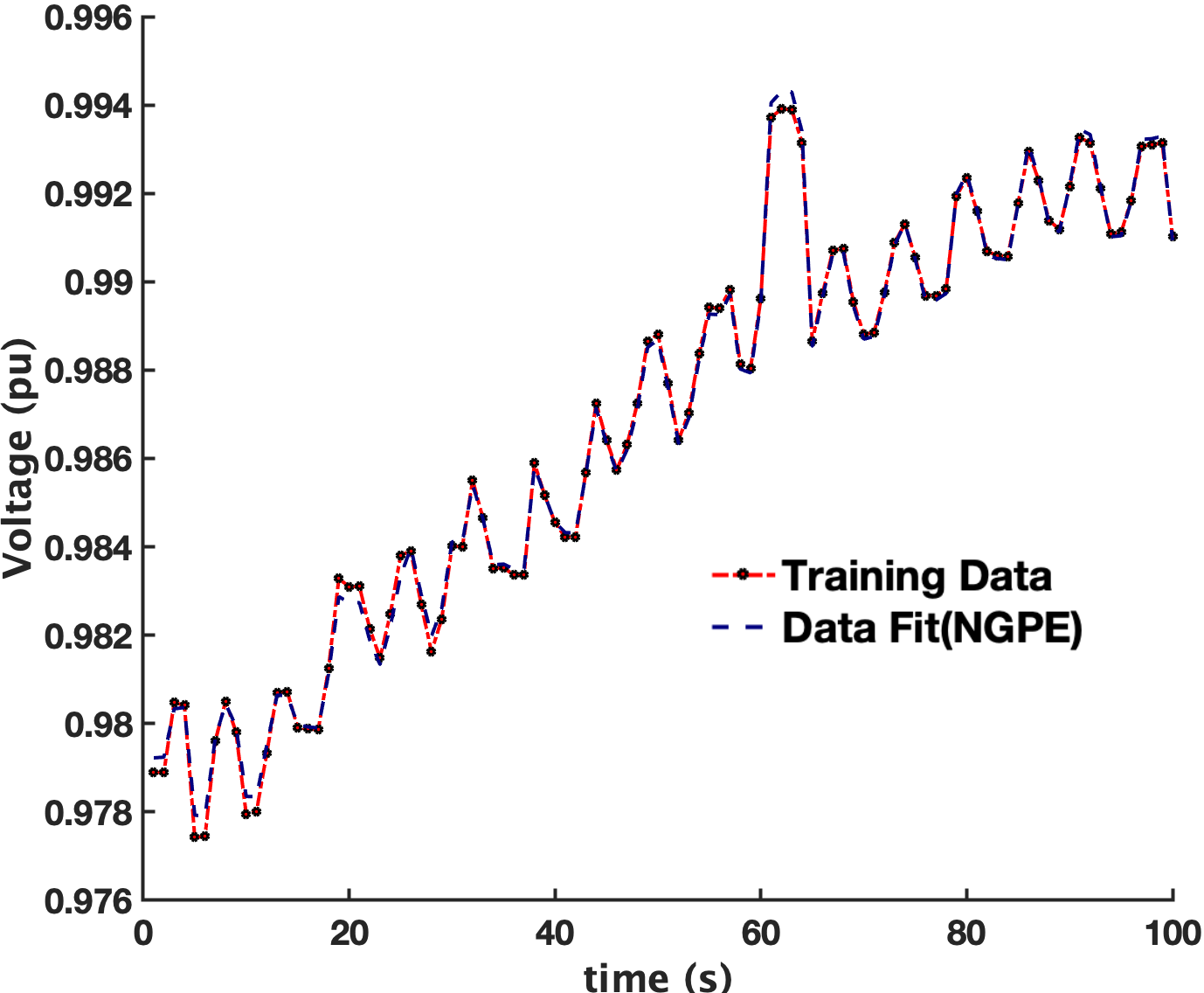} }}%
    \caption{The NGPE results for voltage magnitude at Bus 19; (a) prediction results at the test points; (b) probability density at the test points; (c) fit of through the training data}
    \label{Voltagebus19}
\end{figure}

\begin{figure}%
    \centering
    \subfloat[\centering ]{{\includegraphics[height=3cm,width=9cm]{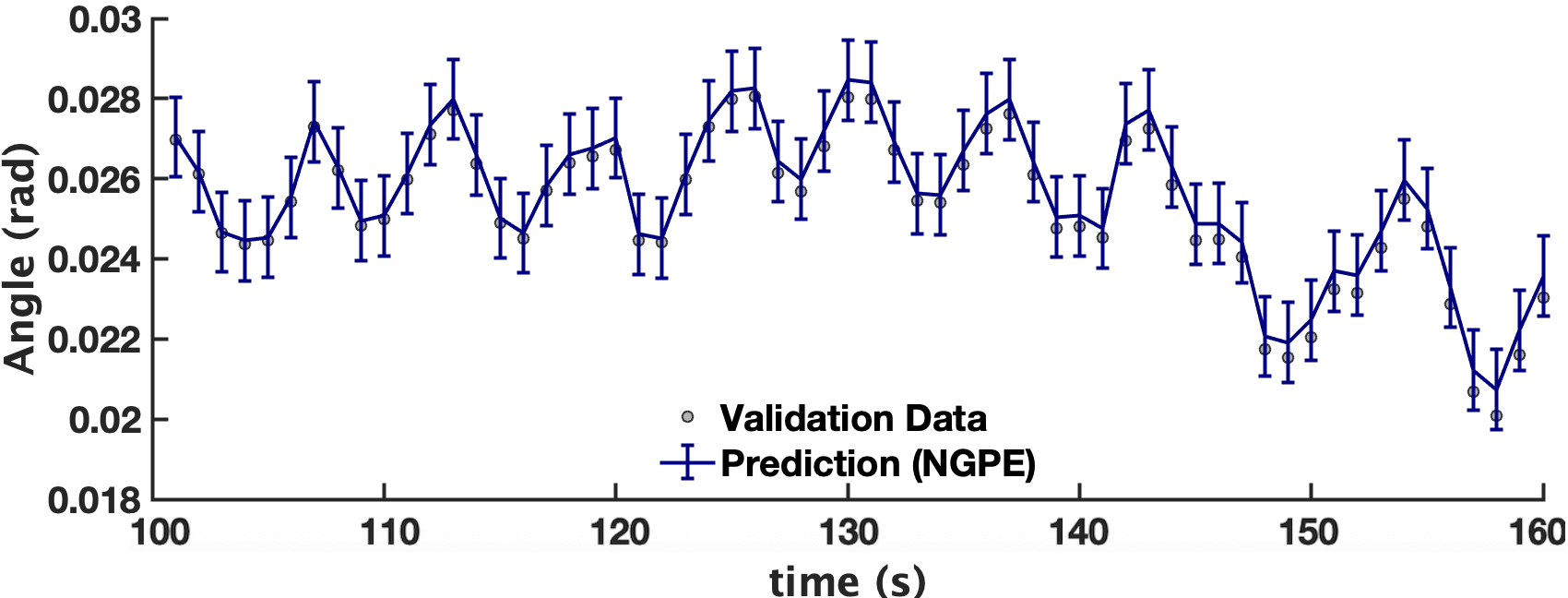}}}%
 \qquad
    \subfloat[\centering  ]{{\includegraphics[height=4.5cm,width=4.5cm]{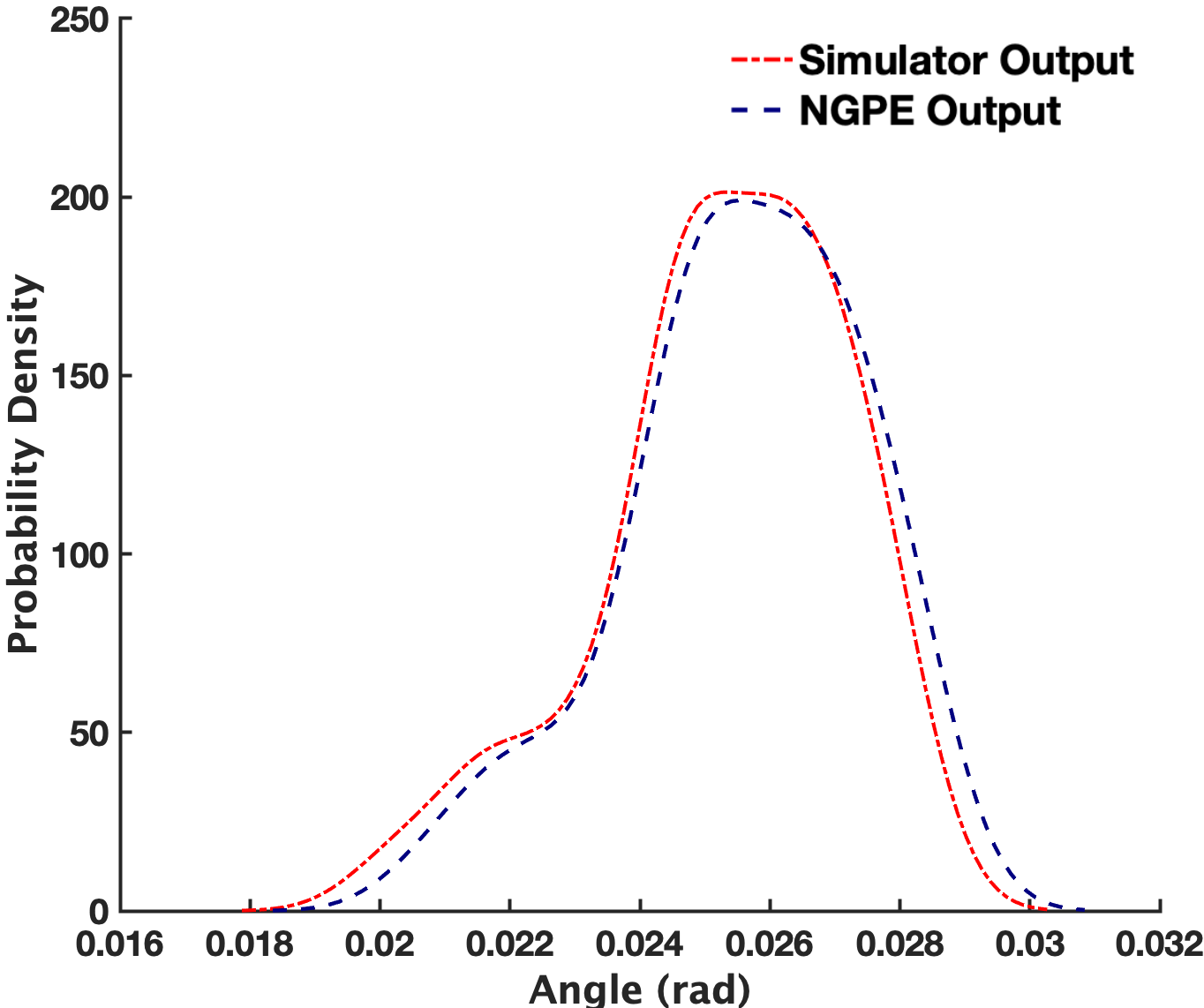}}}%
    \subfloat[\centering ]{{\includegraphics[height=4.5cm,width=4.5cm]{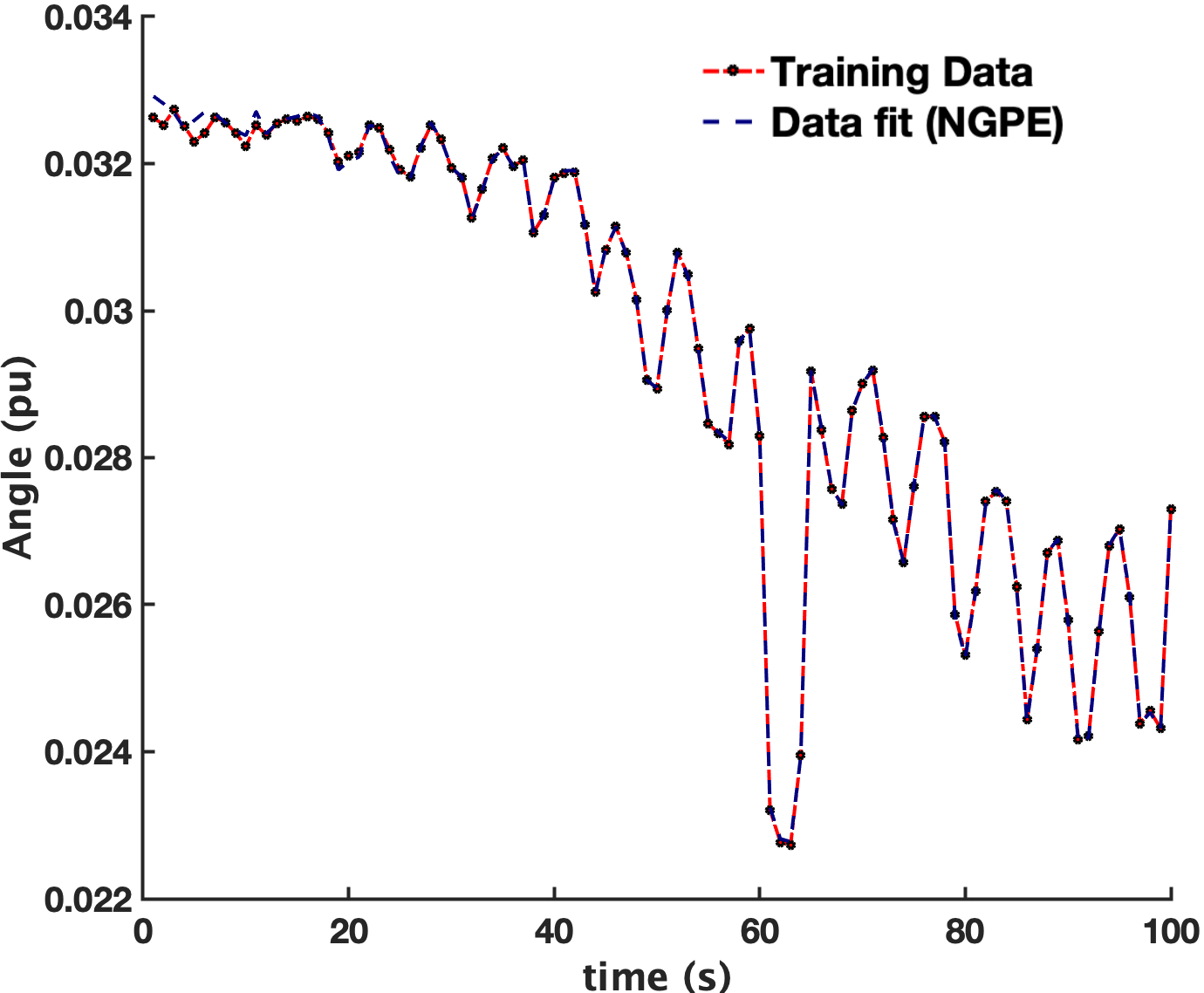} }}%
    \caption{The NGPE results for voltage angle at Bus 19 (a) prediction at the test points; (b) probability density at the test points, (c)fit through the training data}%
    \label{VoltageAngle19}
\end{figure}

\begin{figure}%
    \centering
  \subfloat[\centering ]{{\includegraphics[height=2.5cm,width=9cm]{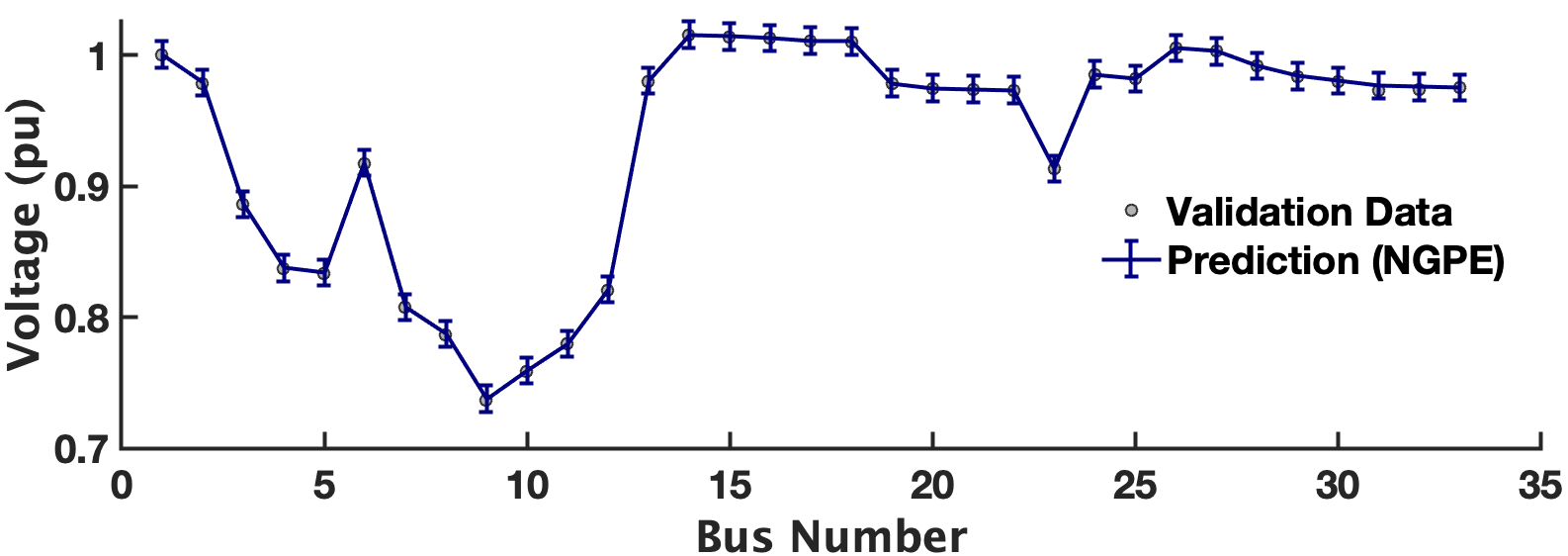}}}%
    \qquad
    \subfloat[\centering ]{{\includegraphics[height=2.5cm,width=9cm]{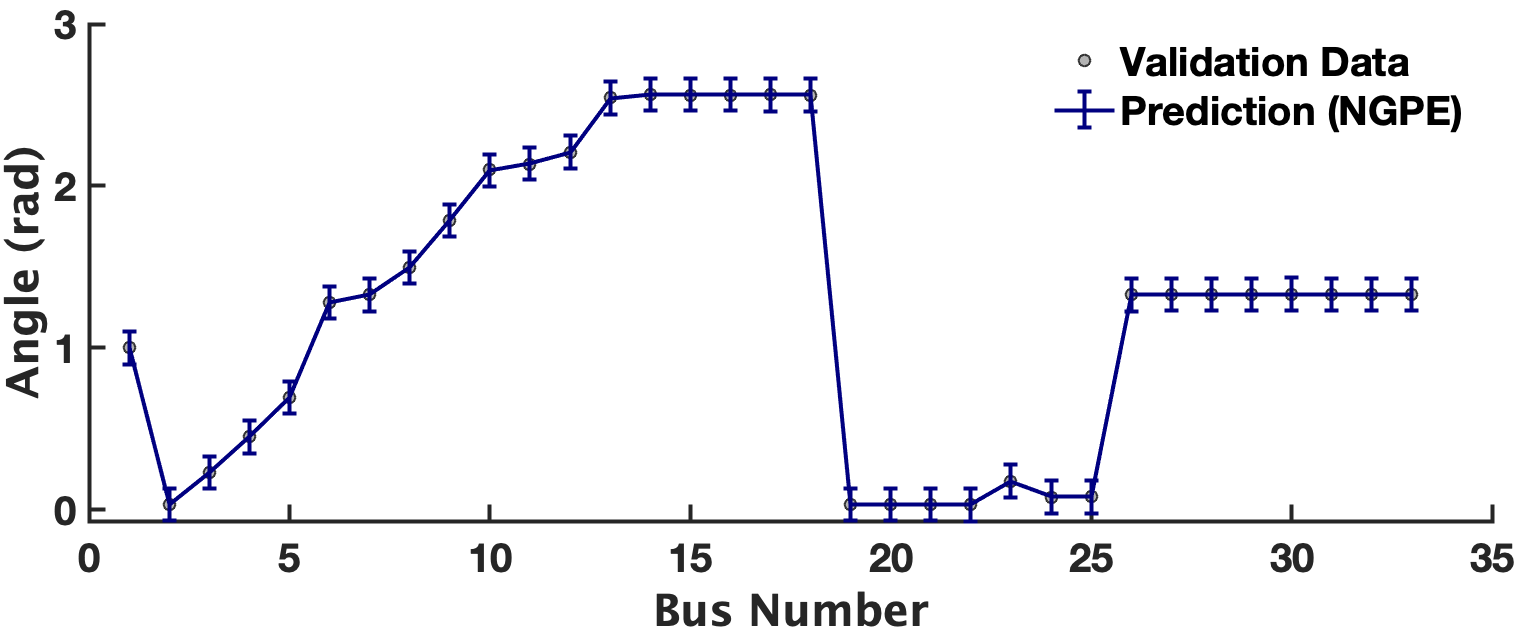}}}%
    \caption{The NGPE predictions of the IEEE 33 bus system with the training data corrupted with $25\%$ outliers; (a) voltage magnitude; (b) angle}%
    \label{outlierL}
\end{figure}

\begin{figure}%
    \centering
  \subfloat{\includegraphics[height=6cm,width=9cm]{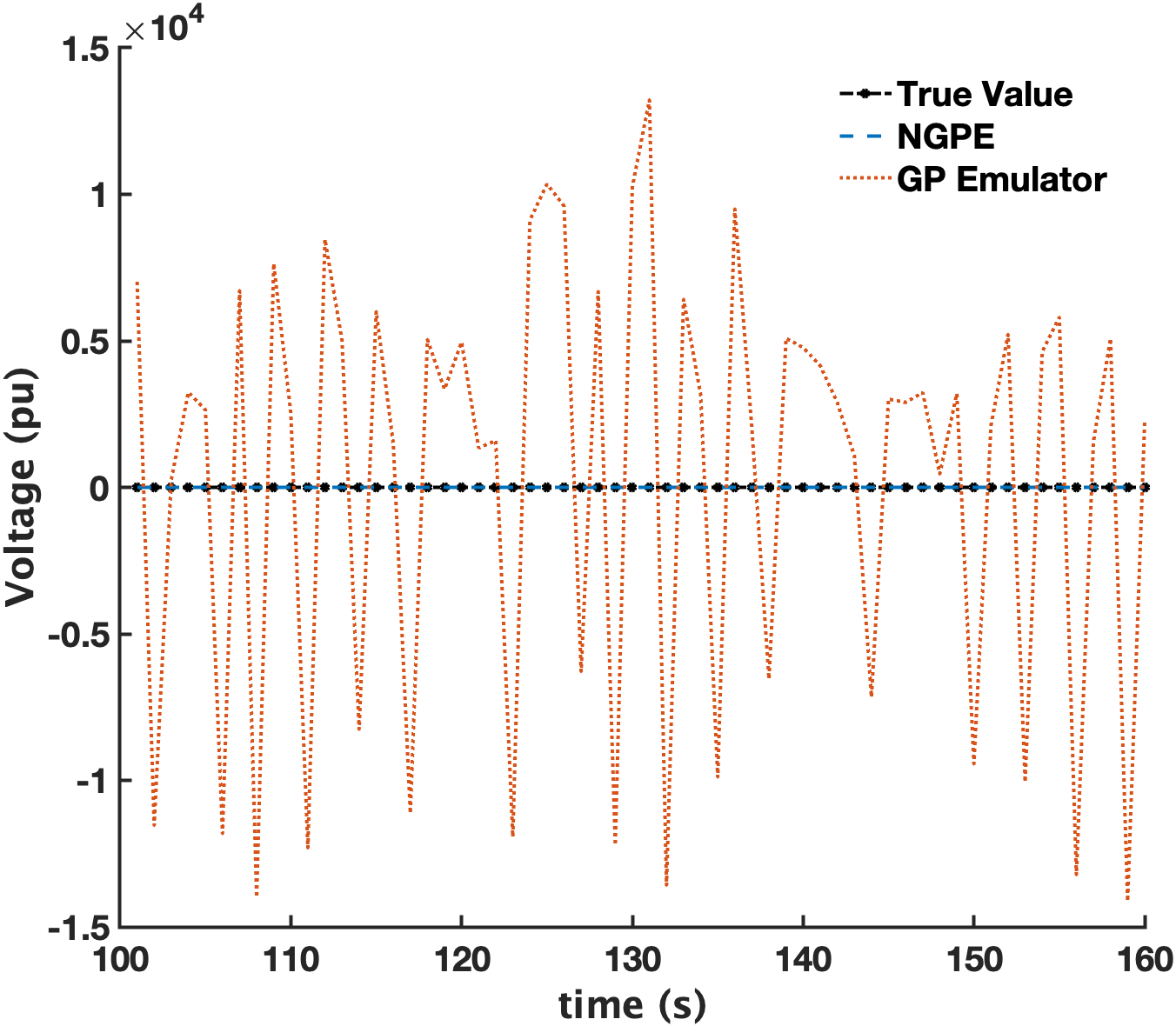}}%
   \qquad
    \subfloat{\includegraphics[height=6cm,width=9cm]{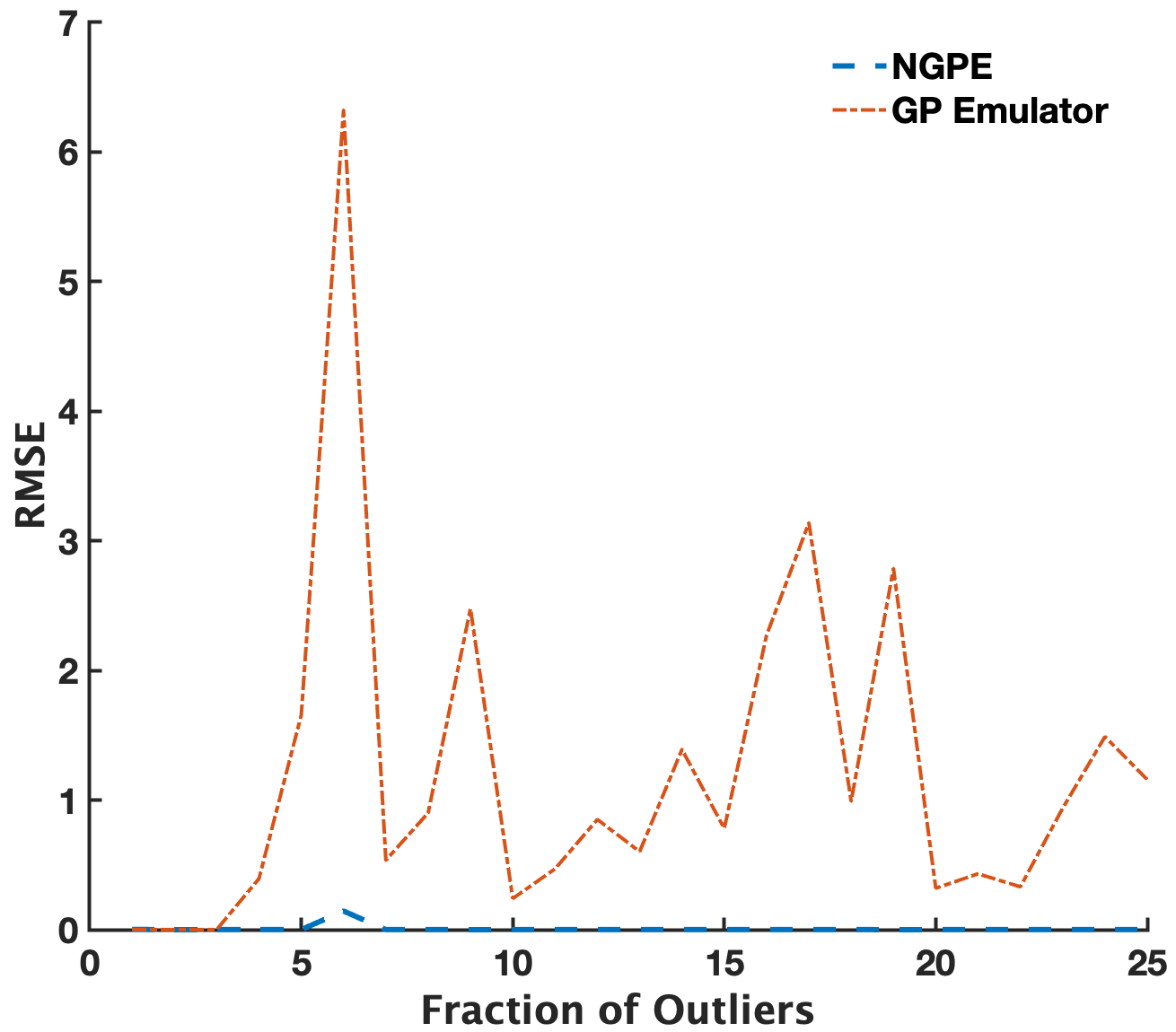}}%
    \caption{The comparison between the NGPE and GP emulator(a)Voltage magnitude at Bus 19;(b) RMSE}
    \label{Comparison}%
\end{figure}

To demonstrate the robustness of the NGPE, we analyze its performance on a standard IEEE 33 bus system with in total four RES, a PV ($P_{G24}$) attached to the Bus 24, and three WGs ($P_{G13},P_{G14},P_{G26}$) are attached to the buses 13, 14, and 26 of capacity 1 kW, 50 kW, 10kW, 10kW, respectively. 
The load shapes over time are considered to be following the Gaussian distributions $P_{t,L}\sim\mathcal{N}(P_{L},0.05P_{L})$ at $t=[1,2,\hdots,N]$. The time-series data considered for the RES power injections are the real measurements with a resolution of 1s. These values for the initial $N=100s$ constitute as the training data and the next $N^{*}=60s$ test points as a validation data. 
To emphasize the effectiveness of the proposed method in presence of outliers, we add the outliers in the measurements $\{P_{G13},P_{G14},P_{G24},P_{G26}\}$ as shown in Fig. \ref{DATA}(a) and to load consumption of the load buses $\{P_{L1},P_{L2},\hdots,P_{L20}\}$ to impose a worst case scenario. 
We observe from Fig.\ref{DATA}(b) that the SHGM estimator downweights the outliers with weights that decrease with the degree of outlyingness. Moreover, it can overcome the masking and smearing effect associated with simultaneous outliers.
The voltage magnitude and angle prediction of the Bus 19 with the fraction of outliers up to $25\%$ in the training data constitute as a benchmark for this study. Fig.\ref{Voltagebus19} and \ref{VoltageAngle19} display the respective results.
We also plot the estimated voltage magnitude and the angle at 33 buses in Fig. \ref{outlierL}.
From the comparison of the results obtained between the NGPE and the GP emulator displayed in Fig. \ref{Comparison}(a) we conclude that because the conventional GP emulator has a non-robust estimation of regression weight vector centered at the weighted least squares estimates, it fails to represent the simulator in presence of outliers. 
Finally, we compare the root mean square errors (RMSE) for the prediction of voltage magnitude at Bus 19 in Fig. \ref{Comparison}(b). We notice, the NGPE offers consistently low RMSE.

\section{Conclusion and future work}
In this paper, we propose a robust methodology based on the SHGM estimator to assess the stochastic dynamics introduced in the power system with high renewable energy penetration. The emulator is trained using finite time series real measurements of the voltage phasors and synchronous generator output power of the IEEE 33-bus distribution system. Due to real data unavailability of the voltage phasors, the latter are obtained through the power flow simulator with real time series measurements of the PV and WGs output power. As a future work, we  will focus on the demonstration of the proposed method on a real-life power distribution system with real voltage phasor measurements. 
\newpage
\bibliography{references1} 
\bibliographystyle{ieeetr}

\end{document}